\documentclass[notitlepage,twocolumn,nobalancelastpage,superscriptaddress,showpacs,aps,pra,10pt]{revtex4-1}
\newsavebox{\mstrut}
\usepackage{soul}
\usepackage{enumerate} 
\usepackage{mathtools}
\usepackage[dvipsnames]{xcolor}
\usepackage{graphicx}
\newsavebox\mysavebox
\usepackage{dcolumn}
\usepackage{bm}
\usepackage{amsmath, amsthm, amssymb}
\usepackage{romannum}
\usepackage{tikz}
\usepackage{amsmath}
\usepackage{subcaption}
\usepackage{hyperref}
\hypersetup{
    colorlinks=true,
    linkcolor=blue,
    filecolor=blue,      
    urlcolor=blue,
    citecolor=blue,
    linkcolor=red,
    breaklinks=true,
}

\begin{document}

\title{All-mirror wavefront division interferometer for Fourier transform spectrometry across multiple spectral ranges}

\author{Ivan Zorin}
\email{ivan.zorin@recendt.at}
\author{Paul Gattinger}%
\affiliation{Research Center for Non-Destructive Testing, Science Park 2, Altenberger Str. 69, 4040 Linz, Austria}

\author{Giovanna Ricchiuti}%
\author{Bernhard Lendl}
\affiliation{%
Institute of Chemical Technologies and Analytics, TU Wien, Getreidemarkt 9/164, 1060 Vienna, Austria
}
\author{Bettina Heise}
\author{Markus Brandstetter}
\affiliation{Research Center for Non-Destructive Testing, Science Park 2, Altenberger Str. 69, 4040 Linz, Austria}
\date{\today}

\begin{abstract}
We report on the design of an all-mirror wavefront-division interferometer capable of spectroscopic studies across multiple spectral ranges\textemdash from the plasma frequencies of metals to terahertz wavelengths and beyond. The proposed method leverages the properties of laser sources with high spatial coherence. A theoretical framework for the interferometer scheme is presented, along with an analytical solution for determining the far-field interference pattern, which is validated through both optical propagation simulations and experimental results. The practical implementation of the spectrometer, using cost-effective off-the-shelf components (knife-edge prisms for separation and recombination), is demonstrated. The system features ultra-broad optical bandwidth, high throughput, simple architecture, dispersion-free operation, and variable arm split ratio. These unique attributes make our approach a prospective alternative to standard Fourier transform spectrometer schemes, specifically tailored to laser-based scenarios. Further, the employed design inherently enables the measurement of the sample's dispersion.
In the experimental section, we demonstrate the feasibility of spectroscopic measurements by coupling the system with a supercontinuum source with more than an octave-spanning range (1.5~\textmu m - 4.4~\textmu m). 
As a proof-of-concept, an experimental demonstration is provided for various applied spectroscopic studies: transmission measurements of polymers (polypropylene) and gas (methane), as well as reflectance measurements of dried pharmaceuticals (insulin products on a metal surface).
\end{abstract}

\maketitle
\section{\label{sec:Intro}Introduction}

Fourier transform (FT) spectrometry is a mature and well-established technique that employs principles of low temporal coherence interferometry for high-resolution spectroscopic measurements\cite{griffiths2007fourier, chalmers_handbook_2001}. Its widespread adoption is attributed to the fundamental features of the FT spectrometer design, which has several distinctive and unique advantages over diffraction and dispersive systems\cite{Saptari2003}. The first advantage is the so-called multiplex sensitivity (Fellgett’s) advantage. It is associated with simultaneous sampling of all spectral channels with a single detector that significantly boosts the signal-to-noise ratio, especially in infrared (IR) non-shot-noise limited systems. Another substantial sensitivity gain is linked to the optical \'etendue (Jacquinot's advantage). In FT spectrometry, the spectral resolution is unconditional on the geometric and optical parameters of the system, such as grating period and slit size. It thus depends solely on the scan length as a reciprocal. Consequently, input slits or apertures, inherent to dispersive systems for achieving spectral resolution, are not required. As a result, FT spectrometers feature high spectral resolution with large optical apertures and, thus, throughput. Furthermore, a standard FT system is backed by a reference laser to linearize the signals while scanning the interferometer arm, leading to a third advantage: prior knowledge of the reference laser's wavelength ensures the high accuracy of time domain measurements and subsequent Fourier domain reconstruction. In state-of-the-art IR FT spectrometers, the spectral response is typically broad and constrained by the detector's spectral response and spectral characteristics of transmission components used in the particular design. For thermal light sources, these limitations dominate, as their spectral range is inherently broad; however, in laser-based spectroscopy, the spectral coverage is often limited by the source bandwidth. Nowadays, most commercial instruments have Michelson's or Mach-Zehnder's configuration and operate in the near- and mid-IR ranges but require either the replacement of a detector or optics to switch between the spectral windows. Technical advantages, affordability and relative simplicity have made FTIR spectroscopic techniques a standard routine tool for chemical analysis in research\cite{DUTTA201773, https://doi.org/10.1002/lpor.202100556,doi:10.1080/10643389.2020.1807450},biomedical\cite{doi:10.1080/05704928.2021.1946822,doi:10.1080/05704928.2018.1431923,doi:10.1080/05704920701829043, sreedhar_high-definition_2015}, and industrial applications\cite{doi:10.1080/19440049.2019.1675909, doi:10.1080/17425247.2020.1737671,doi:10.1080/00387011003601044,PAVLI20181061}.

The main components of any FT spectrometer are a light source, a detector, and an optical interferometer. 
The interferometer's role is to perform the autocorrelation of the source's electric field, enabling the sequential reconstruction of amplitude or phase spectra according to the Wiener-Khinchin theorem.
Concerning the optical scheme, in most scenarios, even for advanced on-chip realizations\cite{doi:10.1177/0003702816638295,fathy_-chip_2020}, a Michelson-type arrangement with a single amplitude divider (such as a plate beam splitter) is used\cite{1176026, doi:https://doi.org/10.1002/0470027320.s0204}. The core interferometer can be, however, advanced mechanically, or its optical part can be extended to other measurement modalities, such as attenuated total reflection, imaging or mapping, using additional components or optics\cite{doi:https://doi.org/10.1002/0470027320.s2602}. Thus, the range of different implementations of FTIR spectroscopic techniques is fairly wide\cite{chalmers_handbook_2001}.

In terms of emitters, conventional scanning FT spectrometry offers versatility by accommodating a wide range of broadband light sources without requiring specialized properties. This primarily includes extended blackbody emitters, which represent the state-of-the-art solution.
Nevertheless, owing to the rapid development of mid-IR photonics, many ultra-bright broadband spatially coherent laser sources have been developed in the last decades, greatly enriching the IR spectroscopy field. Such laser sources are uni-directional, high-power, and provide brightness that is orders of magnitude higher than that of thermal sources\cite{doi:10.1119/1.2344276,SHUKLA201540}. Since the product of brightness and \'etendue is the spectral power that the optical system transmits and employs to probe the sample, mid-IR lasers represent a promising alternative for IR spectroscopy. Among the broad variety of options, supercontinuum sources, characterized by an intrinsic bright and instantaneous ultra-broad emission, are particularly well-suited for FT interferometric methods~\cite{Zorin:22}. Supercontinuum generators are nonlinear laser-driven emitters that feature high laser-like spatial but low temporal coherence (due to the ultra-broad emission profiles)\cite{genty_coherence_2016}, a desired combination for the FT techniques. In particular, recent reports of ultra-broad (covering functional and fingerprint regions and ranges inaccessible to quantum cascade lasers) high-intensity realizations are particularly prospective\cite{Martinez:s,Krebbers:24}.

In this contribution, we aim to re-examine the core of state-of-the-art FT spectrometers\textemdash the interferometer\textemdash in light of the recent development and commercialization of ultra-broadband spatially coherent sources\textemdash table-top sources as opposed to synchrotrons. Thus, we propose a novel and simple optical design of an all-mirror wavefront division interferometer coupled to a mid-IR supercontinuum laser. 
The developed optical system has no transmission optics and is thus dispersion-free and broadly achromatic, with the potential to operate from the ultraviolet to terahertz range. In our design, we exploit the high spatial coherence of the light source to perform lossless division and recombination of the beams. The wavefront division scheme provides a significant improvement over the losses typical for conventional amplitude-division beam splitters. In addition, the selected division method allows the splitting ratio to be freely adjustable, enabling precise dynamic range adjustment in case of low- or high-absorbing samples. 

In the first part (Section~\ref{sec:theory}) of the manuscript, we present the concept of the proposed interferometer as well as a theoretical framework describing optical propagation and formation of interferometric signals, which we relate to experimental data and optical simulations.
In Section~\ref{sec:FTS}, we demonstrate a practical implementation of the proposed system coupled with a mid-IR supercontinuum source and discuss relevant aspects of the design.
In the experimental Section~\ref{sec:Results} we provide proof-of-concept spectroscopic measurements across a range of practically interesting applications. This comprises measurements of vibrational spectra of methane, transmission measurements of polymers, and the detection of low-concentration pharmaceutical residues in reflectance mode.
Since the system is inherently capable of measuring the complex refractive index, this is demonstrated as well.
At last, for completeness, we review and discuss the technical-historical aspects and comparison with previously reported wavefront division schemes in Section~\ref{sec:history}. The conclusion and outlook are presented in Section~\ref{sec:end}.




\section{Wavefront division interferometer: theory and simulations}\label{sec:theory}
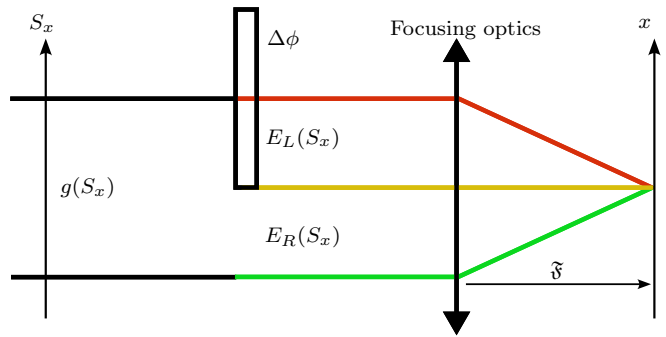
\begin{figure}[ht]
\centering
\begin{tikzpicture}
  \node[anchor=south west,inner sep=0] (image) at (0,0,0) {\includegraphics[width=1\columnwidth]{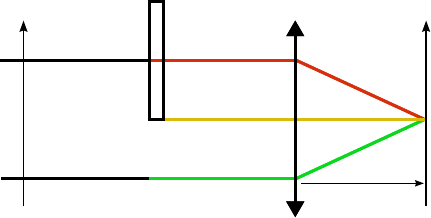}};
  \begin{scope}[x={(image.south east)},y={(image.north west)}]
    \draw (0.42,0.9) node[]{\color{black}\footnotesize $\Delta \phi$};
    \draw (0.84,0.21) node[]{\color{black}\footnotesize $\mathfrak{F}$};
    \draw (0.12,0.45) node[]{\color{black}\footnotesize $g(S_x)$};
    \draw (0.475,0.45+0.15) node[]{\color{black}\footnotesize $E_L(S_x)$};
    \draw (0.475,0.45-0.15) node[]{\color{black}\footnotesize $E_R(S_x)$};
    \draw (0.05,0.95) node[]{\color{black}\footnotesize $S_x$};
    \draw (0.975,0.95) node[]{\color{black}\footnotesize $x$};
    \draw (0.7,0.93) node[]{\color{black}\footnotesize Focusing optics};
  \end{scope}
\end{tikzpicture}
\caption{\label{fig:WD_simple} Linear scheme of the wavefront division interferometer, simplified: the input beam ($g(S_x)$) is wavefront-divided into two parts ($E_L(S_x)$ and $E_R(S_x)$) by a binary rectangular aperture, whereby the phase delay is introduced for the parts independently; the sub-beams combined again into one single beam are focused, creating an interference pattern in the far field. }
\end{figure}

\begin{figure}[hbtp]
\centering
\includegraphics[width=0.5\textwidth]{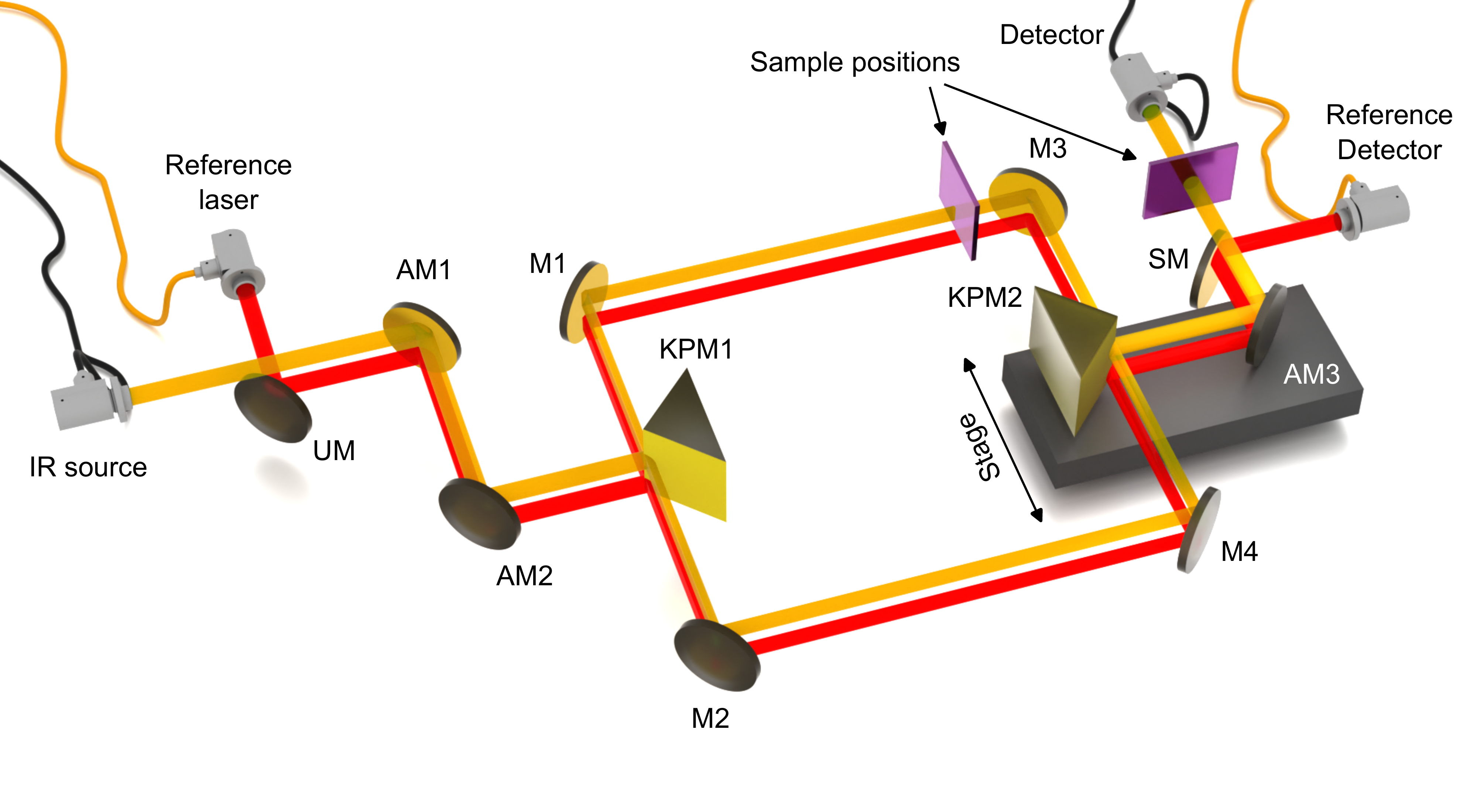}
\caption{\label{fig:WDFTIR} Fourier transform IR spectrometer based on an all-mirror wavefront division interferometer: UM and SM are uniting and separating mirrors respectively coupling two sources to the interferometer at different heights; AM - alignment mirrors; KPM1 and KPM2 are knife-edge prism mirrors dividing and recombining the input wavefront; M1-M4 are flat mirrors of the interferometer arms that form the interferometer together with the KPMs. KPM2 is mounted on a linear stage to vary the length of both arms in time; two possible sample positions are indicated.}
\end{figure}

The basic linearized optical scheme of the proposed wavefront division interferometer concept is shown in Fig.~\ref{fig:WD_simple} (experimental implementation in the FT spectrometer modality\textemdash is displayed in Fig.~\ref{fig:WDFTIR} of the next section). The input beam of the high spatial coherence source $g(S_x)$ is wavefront-divided into two parts $E_L(S_x)$ and $E_R(S_x)$, forming two interferometer arms, respectively. A phase difference ($\Delta \phi$) can be introduced independently for each single arm. The beam after the wavefront divider resembles the input one. The sub-beams produce an interference pattern in the far field of the system, enabling the analysis of the phase and amplitudes of the waves. It has to be noted that the split ratio between the arms can be broadly adjusted by translation of the divider in the lateral dimension ($S_x$-axis in Fig.~\ref{fig:WD_simple}) across the input beam. 

The proposed interferometer, akin to the most well-known Young wavefront division interferometer, exhibits a spatial interference pattern in the lateral dimension. This distinguishes it from the axial modulation (for flat wavefronts) in Michelson amplitude division interferometry. Since the beam after recombination is similar to the input beam with negligible diffraction artifacts, interference is only observed in the focal plane where the paths are indistinguishable\cite{Mandel:91}. 

The interference pattern can be accessed by optical propagation simulations\textemdash in our case performed using an open-source Python toolkit \cite{Vdovin}\textemdash and in analytical form. For simplicity, we consider the analytical form just for one dimension where the wavefront division takes place; the $S_x$ and $x$ coordinates are of relevance for modeling.
The input field can be modeled with a Gaussian intensity profile (a valid approximation for spatially coherent laser sources) and constant phase $\phi$ as:
\begin{equation}
g(S_x) = e^{-S^2_x} e^{i\phi},
\end{equation}
where $S_x$ is the spatial coordinate in the near-field division plane (further $S_y$ is orthogonal to $S_x$).

In the case of a 50/50 split ratio, to which we limit our analytical representation, the act of wavefront division can be expressed as a product of the input field with the Heaviside step function:
\begin{equation}
E_{L/R}(S_x) = g(S_x)H_{L/R}(S_x),
\end{equation}
where $H_{L/R}(S_X)$ are the Heaviside step functions for each half of the beam ($L$ and $R$ indicate "left" and "right" respectively), defined as:
\begin{equation}
H_L(S_x) = \begin{cases}
1 &S_x < 0\\
0 & S_x \geq 0
\end{cases},
\end{equation}

and
\begin{equation}
H_R(S_x) = \begin{cases}
0 &S_x < 0\\
1 & S_x \geq 0
\end{cases}.
\end{equation}

\begin{figure}[hb]
\centering
\begin{subfigure}[b]{0.47\linewidth}
         \centering
         \includegraphics[width=\textwidth]{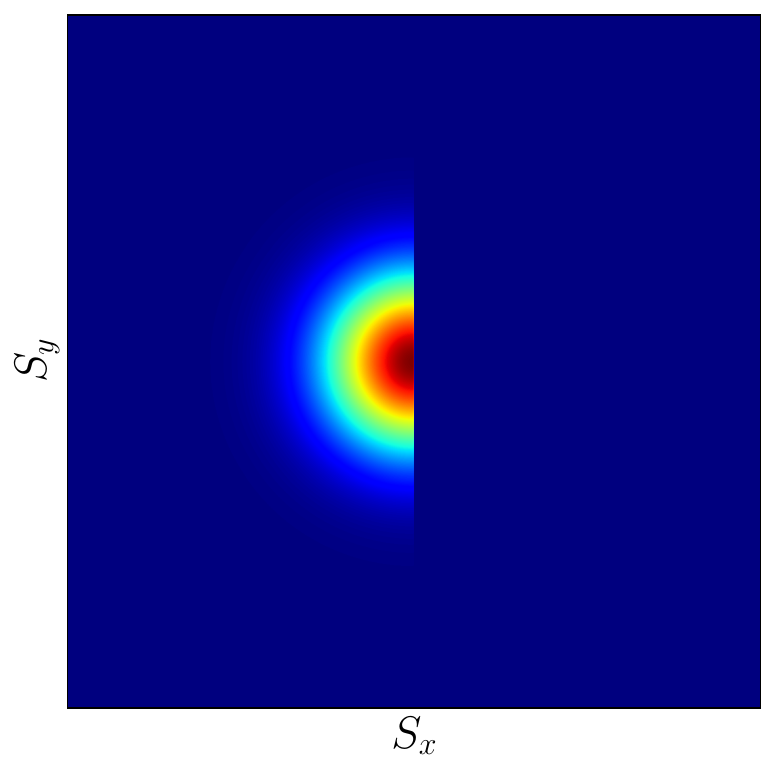}
         \caption{$E_L$ Left beam}
         \label{fig:left_beam}
     \end{subfigure}
     \hspace{5pt}
\begin{subfigure}[b]{0.47\linewidth}
         \centering
         \includegraphics[width=\textwidth]{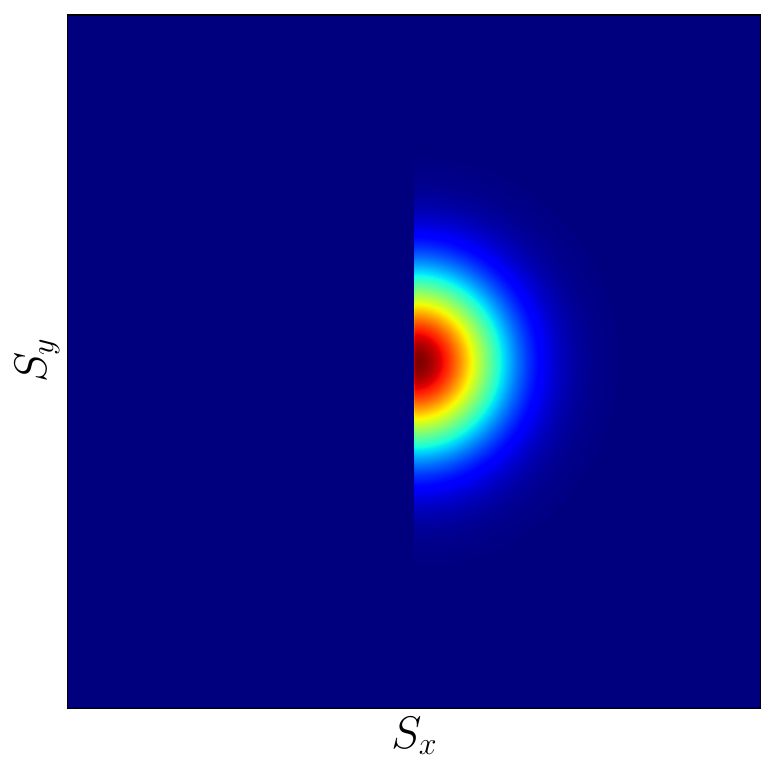}
         \caption{$E_R$ Right beam}
         \label{fig:right_beam}
     \end{subfigure}
     \hfill
\caption{\label{fig:mult_beams} Beam intensity profiles of the interferometer arms after wavefront division; taken at distance 0 (no diffraction).}
\end{figure}

\begin{figure}[hb]
         \centering
         \includegraphics[width=0.47\linewidth]{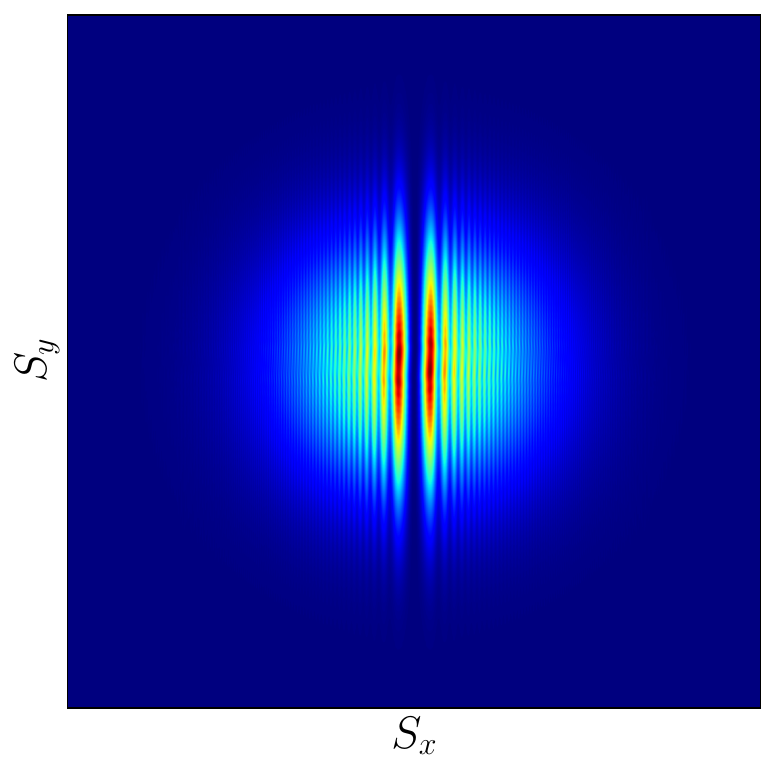}
\caption{\label{fig:diff_beam} Intensity profile of the recombined beam; near-field diffraction is observed.}
\end{figure}

The resulting beams\textemdash snapshots from the optical simulation\textemdash are shown in Fig.~\ref{fig:mult_beams} for illustrative purposes; the beam after recombination showing near-field diffraction is shown in Fig.~\ref{fig:diff_beam}. 

Accessing the interference pattern formed in the focus plane (exemplified for a paraxial positive lens in Fig.~\ref{fig:WD_simple}) requires the use of the Fraunhofer diffraction equation, which can be reduced to a Fourier transform in the far field\cite{Born_Wolf_Bhatia_Clemmow_Gabor_Stokes_Taylor_Wayman_Wilcock_1999}. According to the convolution theorem, the Fourier transform of the product of two functions equals the convolution of the Fourier transforms of those functions. Therefore, in the far field, each of the interfering fields can be represented as a convolution of their corresponding Fourier representations:

\begin{equation}\label{eq:conv1}
E_{L/R}(x) = \mathfrak{F}\{g(S_x)\}*\mathfrak{F}\{H_{L/R}(S_x)\},
\end{equation}
where $x$ is the spatial coordinate in the far field. The Fourier representations of Gaussian functions preserve the Gaussian shape:
\begin{equation}
\mathfrak{F}\{g(S_x)\} = \sqrt{\pi}e^{-x^2}e^{-i\phi},
\end{equation}
for simplicity, we omit the $1/4$ factor in the exponent of the Fourier transform of the Gaussian function.

The Fourier transform of the Heaviside function is given by \cite{bracewell_fourier_2000}:
\begin{equation}
\mathfrak{F}\{H_{L/R}(S_x)\} = \frac{1}{2} \{\delta(x) \pm \frac{i}{\pi x} \},
\end{equation}
where $\delta(x)$ is the Dirac delta function, and the sign depends on whether the step is left- or right-sided.
Thus the convolutions defined in Eqs.~\ref{eq:conv1} take the form of:

\begin{equation}
E_{L/R}(x) = \frac{1}{2} \mathrm{p.v.} \int_{-\infty}^\infty \mathfrak{F}\{g(S_x)\}\left[\delta(x-x') \pm \frac{i}{\pi(x-x')}\right] dx'.
\end{equation}
The first term in the convolution is the convolution of the Gaussian function with the delta function, so it remains unchanged, and the second term has the form of a Hilbert transform. Using the following Hilbert transform pair\cite{King_2009}:
\begin{equation}
f(x) = e^{-x^2} \rightarrow H(f(x)) = -e^{-x^2}\mathrm{erfi}(x),
\end{equation}
where $\mathrm{erfi}(x)$ is the imaginary error function, it is possible to derive the fields as follows:
\begin{equation}\label{eq:field1}
E_{L/R}(x) = \frac{\sqrt{\pi}}{2}\left[ e^{-x^2} e^{-i\phi_{L/R}}\pm ie^{-x^2} e^{-i\phi_{L/R}}\mathrm{erfi}(x)\right].
\end{equation}
The phases $\phi_L$ and $\phi_R$ of the fields are decoupled and can be changed independently.
Thus, the interference pattern in the focal plane can be derived as a coherent sum of incident fields:
\begin{equation}\label{eq:sum_fields}
I(x) = \langle (E_R(x) + E_L(x))(E_R(x) + E_L(x))^* \rangle.
\end{equation}

\begin{figure}[hb]
\centering
\begin{subfigure}[b]{0.47\linewidth}
         \includegraphics[width=\textwidth]{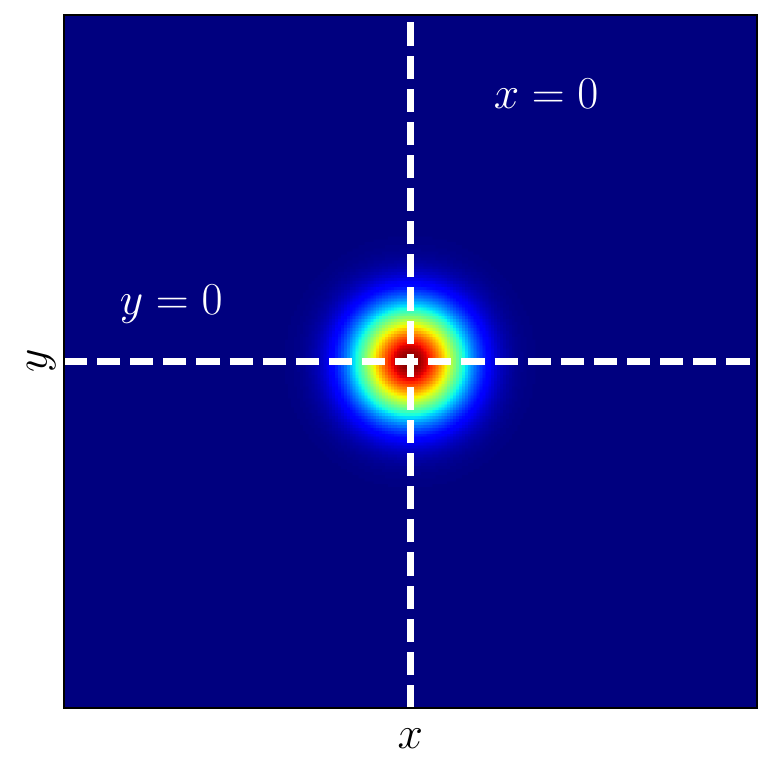}
         \caption{Constructive interference ($\Delta\phi = 0$)}
         \label{fig:const_int}
     \end{subfigure}
          \hspace{5pt}
\begin{subfigure}[b]{0.47\linewidth}
         \includegraphics[width=\textwidth]{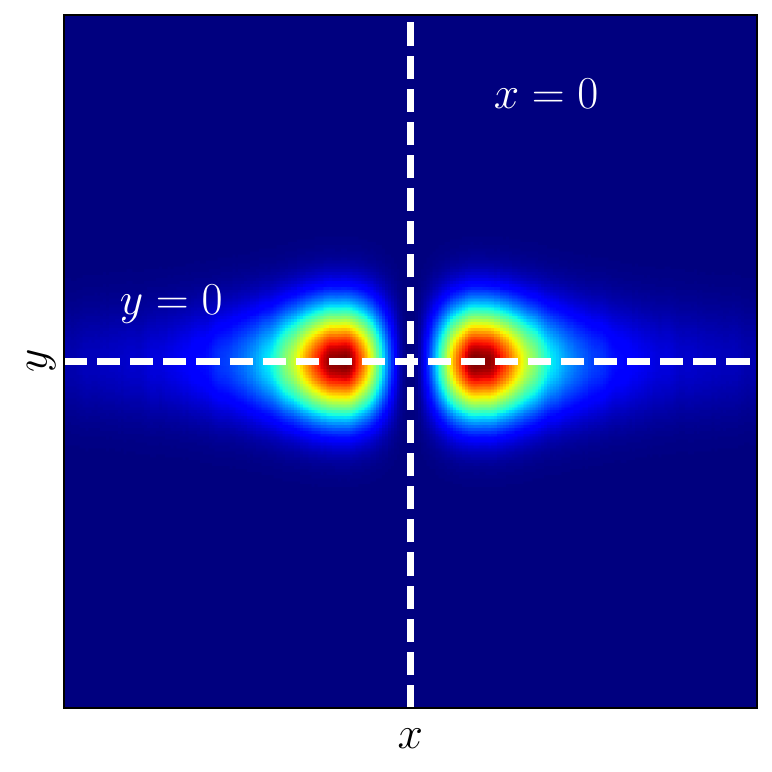}
         \caption{Destructive interference ($\Delta\phi = \pi$)}
         \label{fig:dest_int}
     \end{subfigure}
     \begin{subfigure}[b]{0.47\linewidth}
         \includegraphics[width=\textwidth]{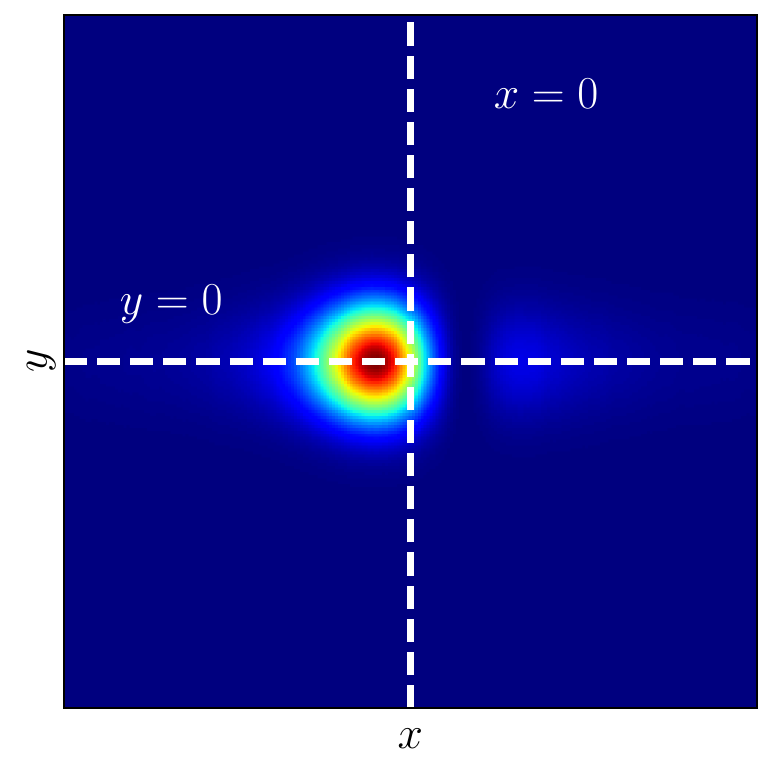}
         \caption{$\Delta\phi = \pi/2$}
         \label{fig:piover2_int}
     \end{subfigure}
          \hspace{5pt}
\begin{subfigure}[b]{0.47\linewidth}
         \includegraphics[width=\textwidth]{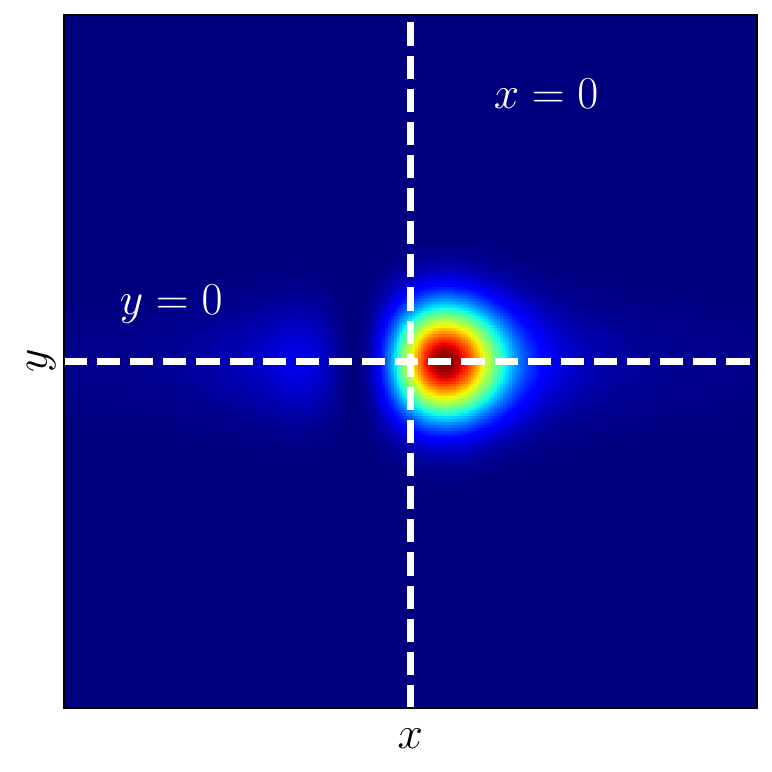}
         \caption{$\Delta\phi = -\pi/2$}
         \label{fig:-piover2_int}
     \end{subfigure}
     \hfill
\caption{\label{fig:interference_patterns} Lateral interference patterns in the far field of a wavefront division interferometer for different representative values of phase delays.}
\end{figure}

Substituting Eqs.~\ref{eq:field1}, Eq.~\ref{eq:sum_fields} takes the following elegant form after meticulous calculations:
\begin{equation}\label{eq:sum_fields_fin}
I(x) = \pi e^{-2x^2} \left[ \cos{\frac{\Delta\phi}{2}} +\mathrm{erfi}(x)\sin{\frac{\Delta\phi}{2}} \right]^2,
\end{equation}
where $\Delta\phi = \phi_R-\phi_L$. It should be noted that the outcome obtained in Eq.~\ref{eq:sum_fields_fin} interestingly resembles the form of a fractional Hilbert transform of a Gaussian function\cite{Lohmann:96,King_2009}.

Equation~\ref{eq:sum_fields_fin} describes the far-field lateral intensity profile and demonstrates the feasibility of using the interferometer for spectroscopic measurements comparable to state-of-the-art schemes, with specific considerations. The solution provides a clear indication of the overall asymmetric behaviour of the interference pattern in the wavefront division interferometer of the proposed type, in contrast to amplitude-division schemes. Thus, the ratio of the pattern scale\textemdash which can be reconfigured by beam scaling or optics adaptation\textemdash to the detector size must be considered during the system design phase. Integration over the full pattern area results in no interference visibility being retrieved. In the case of constructive interference (i.e., $\Delta\phi = 0$), the term containing the imaginary error function turns zero, resulting in a centered Gaussian maximum. In the case of destructive interference (i.e., $\Delta\phi = \pi$), the pattern shows a minimum at the center, on the optical axis ($x=0$, $y=0$), with two pronounced peaks around it. Asymmetric behavior can also be observed in the case of $\Delta\phi = \pi/2$ and $\Delta\phi = -\pi/2$, as the change of sign constructively intensifies one side and suppresses the other. These cases are summarized in Fig.~\ref{fig:interference_patterns}; we omit the scales for $x$ and $y$ axes as they are configuration-dependent and not of principle importance.

It should be noted that the generalized solution for non-50/50 split ratios is quite bulky, as it would contain more terms in the sum defined in~Eq.~\ref{eq:sum_fields}. For non-centered Heaviside functions, the shift in the near-field domain implies additional spatial components in the Fourier domain, so the pattern will retain asymmetry. For this reason, we consider only a special case and propose to use optical simulations for complex split ratios.

\begin{figure}[hbt]
\centering
\begin{tikzpicture}
  \node[anchor=south west,inner sep=0] (image) at (0,0,0) {\includegraphics[width=1\linewidth]{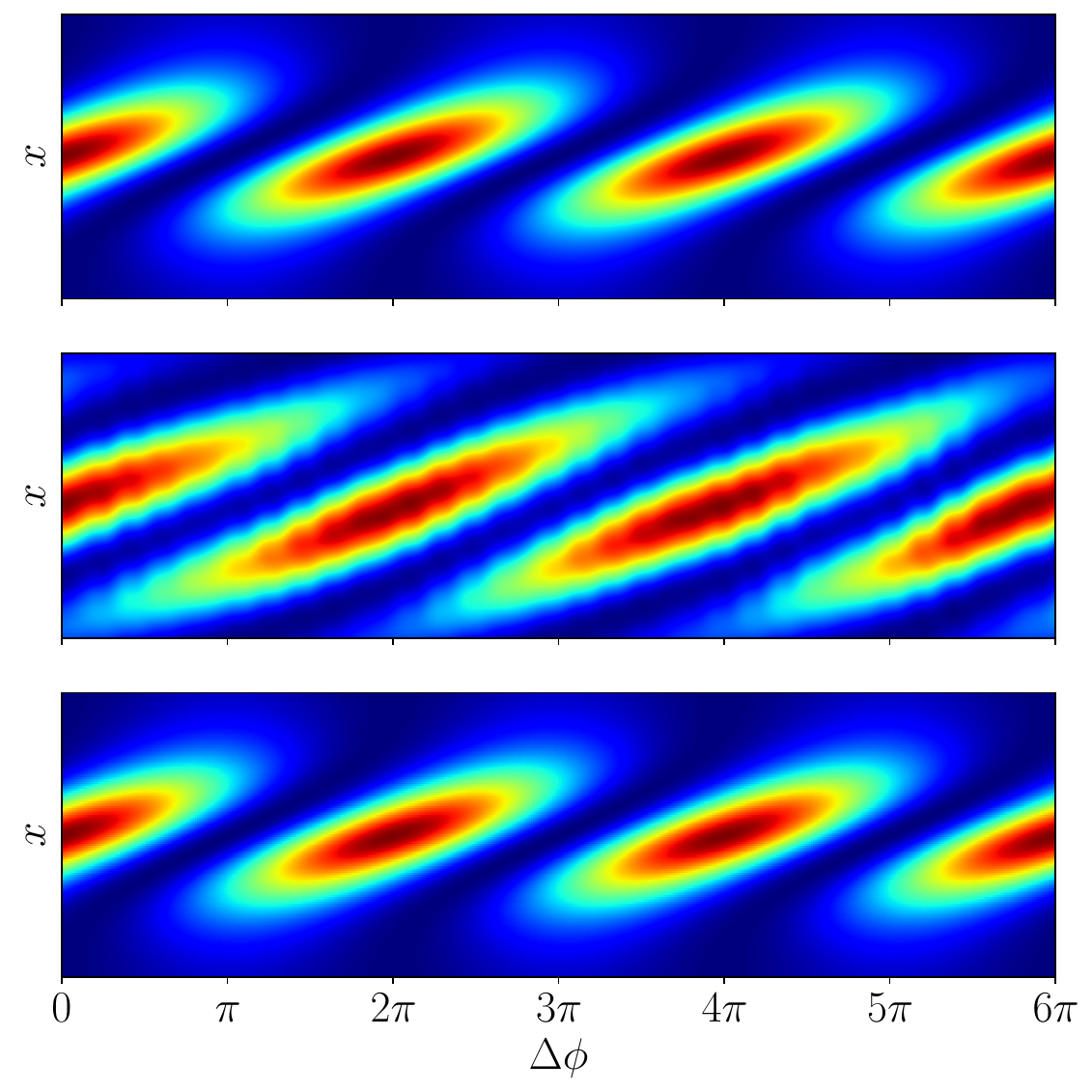}};
  \begin{scope}[x={(image.south east)},y={(image.north west)}]
    \draw (0.27,0.95) node[]{\color{white}\footnotesize (a) Analytical solution};
    \draw (0.22,0.65) node[]{\color{white}\footnotesize (b) Experiment};
    \draw (0.27,0.33) node[]{\color{white}\footnotesize (c) Optical simulation};  \end{scope}
\end{tikzpicture}
\caption{\label{fig:int_comparison} Far-field interference patterns in the wavefront division interferometer: (a)~analytical solution, (b)~experimental data, and (c)~optical simulations; displayed for $y=0$ (wavefront division plane, centered for $x$), the plane corresponds to the $y=0$ white dashed line in Fig.~\ref{fig:interference_patterns}; for the experimental data, the x-axis has a size of 550~\textmu m (100~px), the pixelated structure is due to the limited temporal resolution of the camera during fast phase scanning.}
\end{figure}
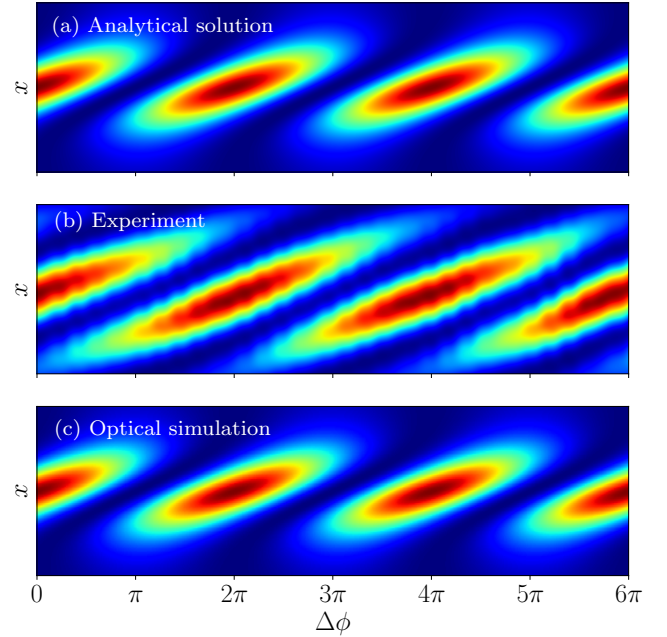

In order to verify the analytical solution and the results obtained with optical propagation simulations, we evaluated the far-field interference patterns experimentally. For this purpose, the optical arrangement of the wavefront division interferometer, discussed in detail in Section~\ref{sec:FTS}, was used. A continuous-wave monochromatic helium-neon laser (632.8 nm) was used to obtain a clear interference pattern. After beam recombination, the far-field image was acquired using a lens with a focal length of 400~mm. The long focal length was used to ensure that the camera (Ximea, xiQ, MQ022RG-CM, 5.5~\textmu m pixel size, operated at 700~fps) had enough pixels to resolve the image. The phase $\phi$ has been scanned using a linear translational stage.

The experimental data, matched with the results of the simulations\cite{Vdovin} and the analytical solution (obtained using Eq.~\ref{eq:sum_fields_fin}), are shown in Fig.~\ref{fig:int_comparison}. The results are displayed for the $x\Delta\phi$ plane (spatial coordinate in the middle of the beam [$y=0$] and phase delay). The findings are in good agreement. The overall asymmetric behaviour is well distinguished. The slight deviation between the experimental and computed data can be attributed to astigmatism, different initial beam profile and, for example, camera positioning not exactly in the focus.

The results obtained in this section establish the theoretical basis for the interferometer design. The gained outcomes also have practical significance. For example, for point detector measurements, it can be seen that the size of the detector should be smaller than the size of the entire interference pattern. Otherwise, integration over the pattern area ($xy$ plane) leads to smearing out of the integrated interference visibility and, hence, to the reduction of the modulation depth.

\section{Wavefront division Fourier transform IR spectrometer}\label{sec:FTS}

The practical implementation of an all-mirror FT spectrometer based on the introduced principles of wavefront division interferometry is shown in Fig.~\ref{fig:WDFTIR}. 
As in a conventional amplitude division FT spectrometer, the system employs two light sources: a broadband IR source (mid-IR supercontinuum source with high spatial coherence) and a quasi-monochromatic reference laser (continuous wave helium-neon laser). The employed supercontinuum source (NKT Photonics, SuperK, ZrF\textsubscript{4}-BaF\textsubscript{2}-LaF\textsubscript{3}-AlF\textsubscript{3}-NaF fiber based) provides a broadband emission coverage in the range from 1.1~\textmu m to 4.4~\textmu m (the near-IR part containing a strong pump line at 1.55~\textmu m is excluded using a spectral filter). The system operates at a 2.5~MHz repetition rate; the average power is about 490~mW. The beam quality M\textsuperscript{2} is 1.09. The reference laser is deployed to produce a distinct high-quality sinusoidal interferogram, which is utilized to linearize the low-coherence IR interferogram during the mechanical scanning of the arms. The beams are co-aligned to enter the interferometer at different heights using a uniting mirror (UM), i.e. they are collinear along the optical axis. The size of the optical components allows them to be handled separately.
 \begin{figure*}[hbt]
\begin{subfigure}[b]{0.48\linewidth}
         \centering
         \includegraphics[width=1\columnwidth]{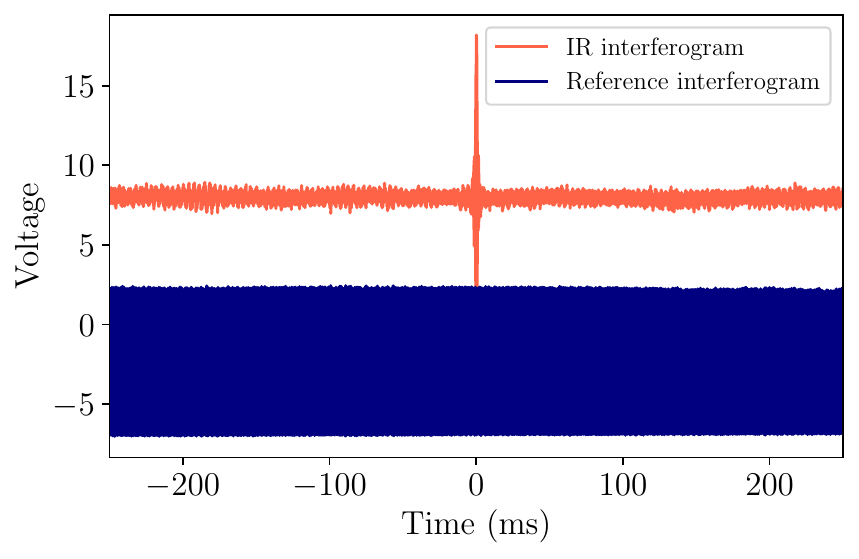}
         \caption{Raw interferograms: 500~ms base (10~mm/s stage speed)}
         \label{fig:inte2}
     \end{subfigure}
\begin{subfigure}[b]{0.48\linewidth}
         \centering
         \includegraphics[width=1\columnwidth]{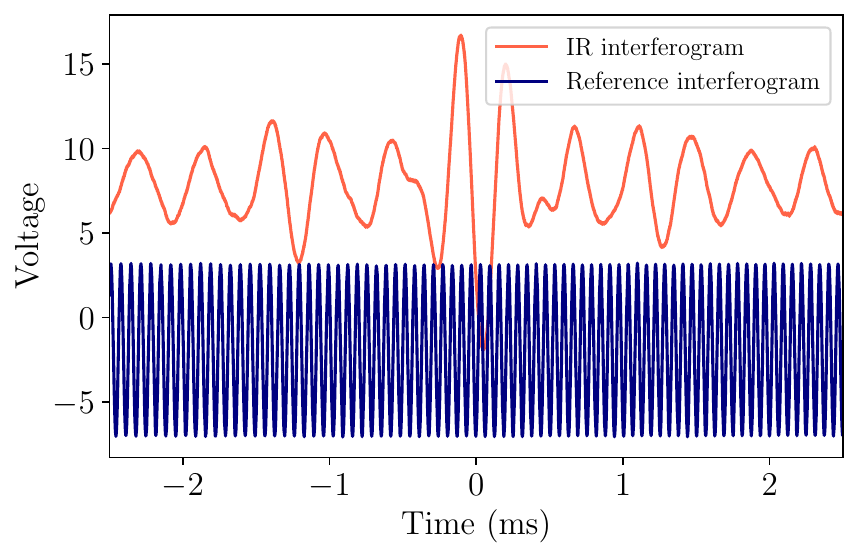}
         \caption{Raw interferograms: zoomed-in, 5~ms base}
         \label{fig:inte1}
     \end{subfigure}
\caption{\label{fig:raw_ints} Typical raw interferograms captured using the developed Fourier transform spectrometer: a low-coherence IR interferogram (obtained for the broadband IR supercontinuum source) and the reference signal (obtained for the quasi-monochromatic helium-neon laser).}
\end{figure*}

The beams are guided normally to the first knife-edge right-angle gold prism (KPM1), which divides the wavefront and thus forms the arms of the interferometer. The two halves of the input fields (assuming a 50/50 split ratio) are directed by the interferometer mirrors (M1-M4) to another identical prism, KPM2, where the combination takes place. Hence, the interferometer resembles a Mach-Zehnder type configuration with a symmetric right-angle arrangement formed by four mirrors. The IR and reference beams are separated the same way they were combined by a separating mirror (SM). The interference pattern is recorded in the far field using corresponding single-point detectors. In the experimental setup, the light is coupled into a single-mode optical fiber to facilitate connection to the detector and, e.g., to a reflection measurements head (in case of reflection measurements). Off-axis parabolic mirror couplers are used so that the far-field image is formed directly in the plane of the fiber cores. A small longitudinal translation of KPM2 (perpendicular to the path length scanning) enables\textemdash akin to induced astigmatism\textemdash scaling of the interference pattern in the focal plane to achieve high interference visibility (integrating over the core area). The theoretical size of the interference pattern for the experimental system is 21~\textmu m; the core size used for spatial filtering (can be replaced by an aperture) is 9 ± 0.5~\textmu m. This is a direct consequence of the theoretical considerations discussed in the previous section. A simple pinhole with properly selected optics (to not sacrifice the throughput) or a single-point detector with long-focus mirrors can be used for the same purpose. In the developed spectrometer, the KPM1 was aligned using a manual stage to maximize the amplitude of the central burst. 
The highest interference visibility\textemdash the actual signal measured in FTIR\textemdash can be observed when the amplitudes of the electric fields in both arms of an interferometer are equal. If an absorbing sample is placed in one arm of the interferometer (e.g. in dispersion measurements), the arms will unbalance, reducing the signal. This can be compensated by introducing a non-equal split ratio, that can be dynamically changed in the broad range.

In order to emphasize the broadband nature of the designed spectrometer, a novel fast mid-IR pyroelectric detector (Laser Components, PR No.1) capable of operation from UV to THz range has been employed. The IR detector has an ultra-thin LiTaO\textsubscript{3} pyroelectric layer (6~\textmu m) thus providing a high responsivity of 70000~V/W (typical) and a detectivity of $2\times10^8$~$\mathrm{cm}\cdot\mathrm{\sqrt{Hz}/\mathrm{W}}$ (at 1~kHz). The bandwidth of the detector is extended up to 100~kHz\cite{app14103967}. Since pyroelectrics are sensitive to alternating signals, i.e., temperature changes, the system is well suited to the principles of FT spectrometry. The modulation induced by interference (time-varying component) is detected directly. Thus, the IR interferograms are captured and mean-centered at zero, with the constant component excluded inherently. Since the detector is slower than photonic detectors, the pulses of the source are naturally integrated, and no demodulation is required. It should be noted that despite the high sensitivity, the pyroelectric detector has a limited dynamic range. The dynamic range, characterized in the time domain under the given FT spectrometry measurement parameters, is approximately 740 (ratio of center burst amplitude to standard deviation of the noise level)  and is constrained by the fixed amplification of the detector electronics.
The light intensity is controlled in front of the sensitive area to avoid oversaturation. The reference interferograms were recorded using a visible photo receiver (Femto HCA-S).

Arm length differences are scanned by lateral scanning of the KPM2 using a motorized linear stage (Zaber, X-LHM050A-E03). It should be noted that the KPM2 scanning simultaneously varies the length of both arms, effectively doubling the scanning range similar to standard FTIR spectrometers. An alignment mirror AM3 is fixed on the same stage to ensure pointing stability in subsequent optical components of the system.

The signals in the implemented FT spectrometer were recorded using an oscilloscope (Teledyne Lecroy, HDO6104A, 1~GHz bandwidth, 10~GS/s). One channel is allocated to the IR interferogram and the other to the reference interferogram. The post-processing of the recorded interferograms was implemented in the Python environment. The developed algorithm involved kernel linearization, apodization (Blackman–Harris window), zero padding, Fourier transform, and Mertz correction.
Figure~\ref{fig:raw_ints} displays raw unprocessed wavefront division interferograms (the reference signal is offset by 7.5~V for clarity), recorded with the oscilloscope. 

There are two possible sample locations in the system. For IR transmission measurements, the sample can be placed either in the combined beam or in one of the interferometer's arms. The first case resembles the standard scenario of absorption measurements in conventional systems. The latter case is of particular interest because the wavelength-specific phase delay (dispersion) introduced by the sample remains uncompensated and hence enables complex refractive index measurements. IR dispersion spectroscopy is of major applied relevance due to the linear dependence of the refractive index on the concentration\textemdash beyond the validity of the Bouguer–Beer–Lambert law\cite{DABROWSKA2023122014, https://doi.org/10.1002/cphc.202000018,doi:10.1177/00037028241258109}.
On the other hand, positioning the sample in the combined beam is reasonable, for example, for reflectance or attenuated total reflection measurements, since no change in the interferometric arm scheme is required.

\section{Proof-of-concept measurements}\label{sec:Results}

\begin{figure}[h]
\centering
\begin{tikzpicture}
  \node[anchor=south west,inner sep=0] (image) at (0,0,0) {\includegraphics[width=1\columnwidth]{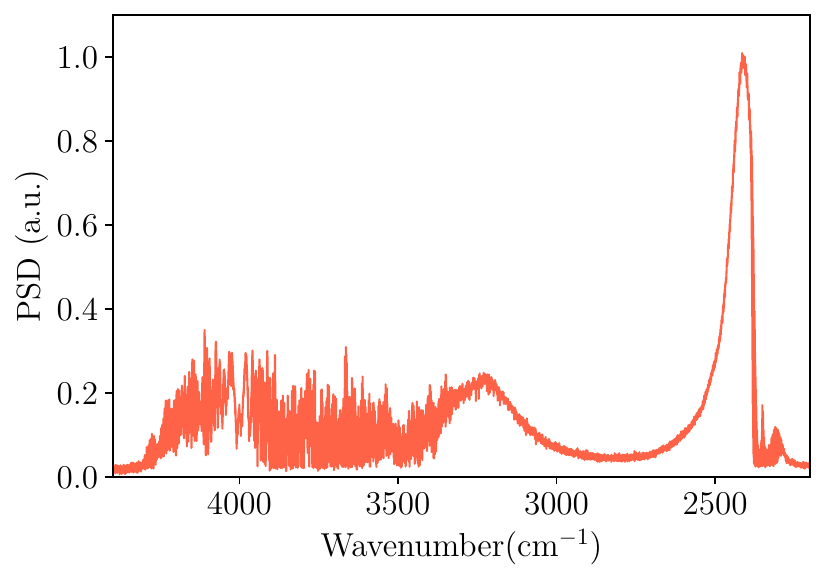}};
  \begin{scope}[x={(image.south east)},y={(image.north west)}]
     \draw (0.5,0.68) node[]{\includegraphics[width=0.4\columnwidth]{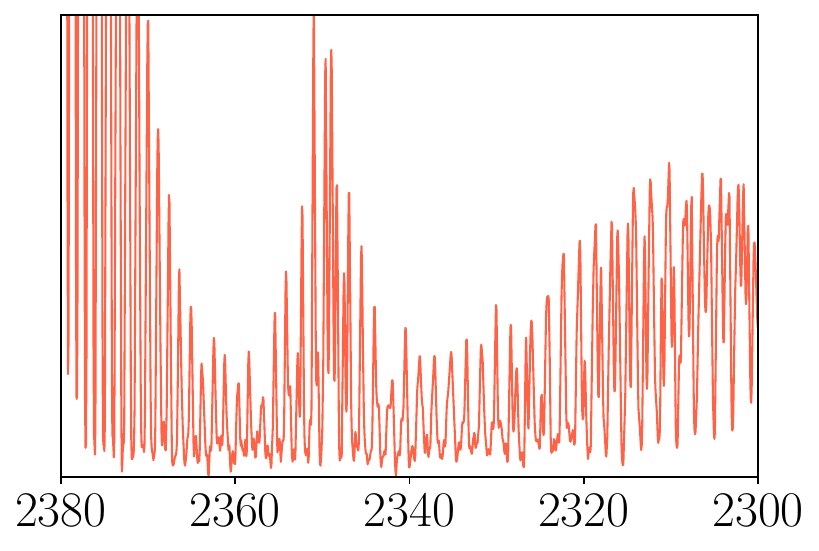}};
  \end{scope}
\end{tikzpicture}
\caption{\label{fig:SC_emission}Normalized emission power spectral density of the IR supercontinuum source measured using the developed wavefront division spectrometer at a spectral resolution of 0.2~cm\textsuperscript{-1}; due to the high resolution, the sharp absorption of ambient water vapour is observed, the inset depicts the magnified absorption band structure of ambient CO\textsubscript{2}.}
\end{figure}

In this section, IR spectroscopic proof-of-concept measurements performed with the experimental FT spectrometer are presented. An extensive spectroscopic analysis is not provided; different application scenarios common for IR spectroscopy are demonstrated to display the concept's feasibility. Since the signals recorded with the detector are identical to ones recorded with state-of-the-art Michelson-type FT spectrometers, the accuracy, stability, dynamic range and resolution are purely engineering aspects related to the particular configuration of the system. The capabilities of the implementation are demonstrated in several applied scenarios. Prior to demonstrating the spectroscopic measurements, in Fig.~\ref{fig:SC_emission} we display the power spectral density of the used supercontinuum source recorded with the developed FT spectrometer. The total scanned arm length difference for this measurement is 5~cm, so the achieved spectral resolution is 0.2~cm\textsuperscript{-1}, comparable to standard table-top commercial systems.
The measurements presented in this section are performed at different spectral resolutions but at the fixed stage scan speed of 12~mm/s; for optical path difference scanning, it is effectively doubled to 24~mm/s. Thus, for instance, for 0.4~cm\textsuperscript{-1} spectral resolution, a single measurement takes around 1~sec. The spectra reconstruction algorithms have been implemented in Python and made open access\cite{WDFTgit}.

\subsection{Transmission measurements: gas}
In order to illustrate the capability of the developed spectrometer for high-resolution detection and analysis of gases, transmission measurements of methane (CH\textsubscript{4}) were performed. 
The absorption measurements were carried out in the combined beam configuration.
As a sample, a 10~cm long sealed gas cell (Wavelength References) with 100~Torr CH\textsubscript{4} (balanced with N\textsubscript{2} to 500~Torr) was used. 

Methane exhibits strong and fine spectral features in the mid-IR range at around 3019~cm\textsuperscript{-1} attributed to vibrations of C-H (fundamental triply degenerate asymmetric stretching)\cite{doi:10.1098/rspa.1952.0109}. 
Since methane is a greenhouse gas, as well as an important combustion-related species, the measurements also emphasize practical potential.

\begin{figure}[ht]
         \centering
         \includegraphics[width=1\linewidth]{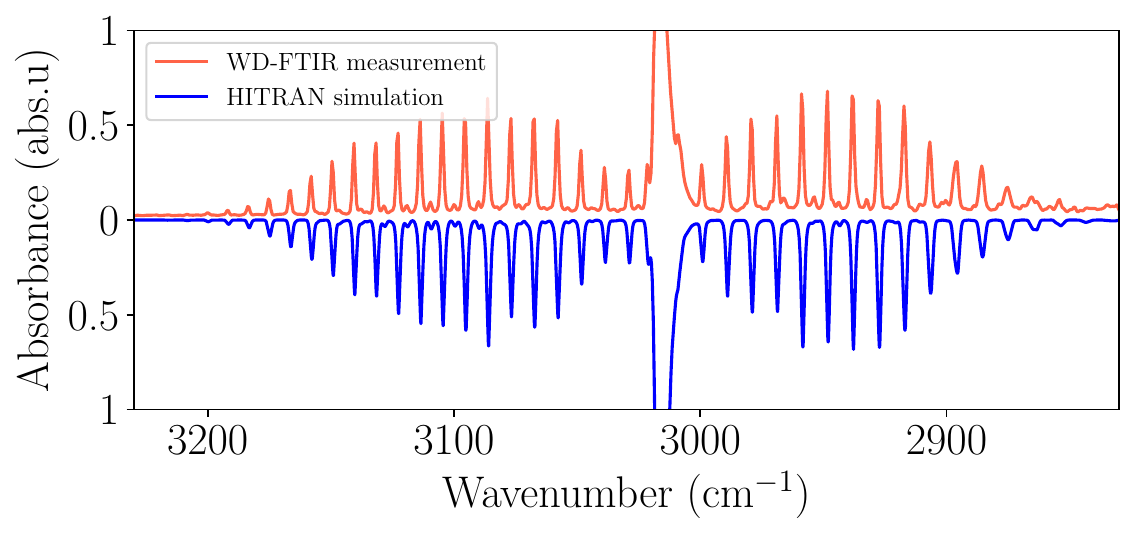}
\caption{\label{fig:ch4-abs} Absorbance spectra of methane obtained with the developed Fourier transform wavefront division spectrometer (WD-FTIR); the blue line, shown for illustrative purposes, represents a HITRAN database simulation as a reference.}
\end{figure}

Figure~\ref{fig:ch4-abs} depicts an absorbance spectrum calculated according to the Bouguer–Beer–Lambert law (decadic form) as:
\begin{equation}
A(\tilde{\nu}) = -\log10\left[\frac{I(\tilde{\nu})}{I_0(\tilde{\nu})}\right],
\end{equation}
where $I$ and $I_0$ are the intensities of transmitted and incident radiation, respectively, and $\tilde{\nu}$ is the wavenumber.
The measurement is performed at around 1.4~cm\textsuperscript{-1} resolution. A higher resolution is feasible (as demonstrated before), however, in this case, strong absorption of highly concentrated gas causes total absorption due to the limited dynamic range of the detector.

Figure~\ref{fig:ch4-abs} also displays an absorbance spectrum simulated using the HITRAN spectroscopic database\cite{GORDON2022107949} for the exact given cell parameters. It should also be noted that the absorption in the Q branch is strong, leading to total absorption. For this reason, the plots are constrained and focus primarily on the P and R branches. 

\subsection{Complex refractive index measurements: polypropylene}

In this subsection, measurements of the complex refractive index for a solid sample, a thin polymer film, are demonstrated and exemplified. As a sample, a 6~\textmu m thick polypropylene film was used (Chemplex Industries, inc., Cat.No.425). 

\begin{figure}[hbt]
\begin{subfigure}[b]{1\linewidth}
         \centering
         \includegraphics[width=\textwidth]{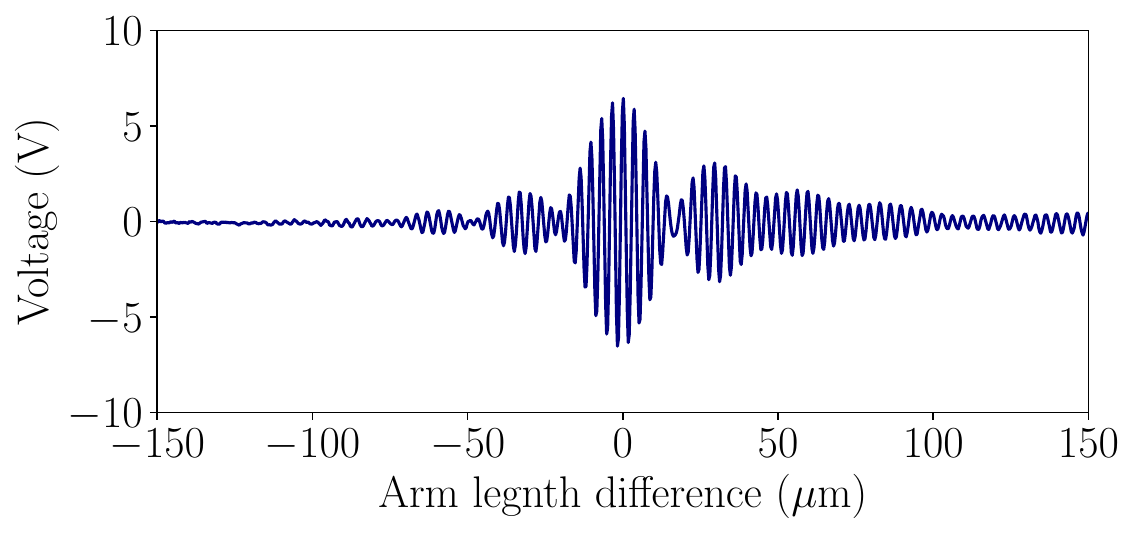}
         \caption{IR interferogram with a polypropylene film inserted}
         \label{fig:pp_int_sa}
         \end{subfigure}
\begin{subfigure}[b]{1\linewidth}
         \centering
         \includegraphics[width=\textwidth]{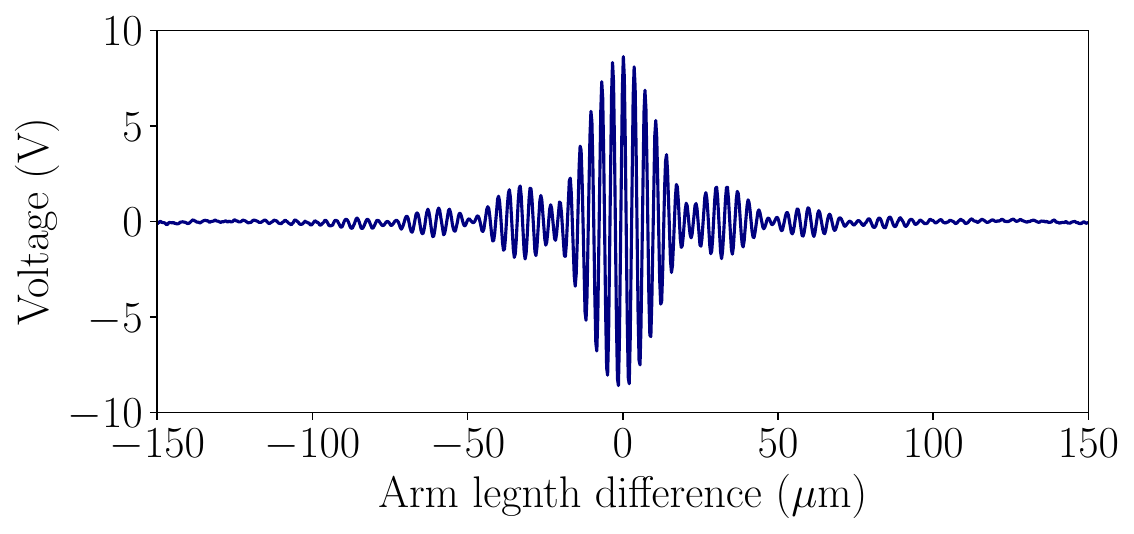}
         \caption{Dispersion-free IR interferogram (sample not inserted)}
         \label{fig:pp_int_ref}
     \end{subfigure}
\caption{\label{fig:pp-ints} IR interferograms for the reference measurement (dispersion-free, no sample insertion) and the complex refractive index measurement (the polypropylene film is in one of the arms of the interferometer).}
\end{figure}

The sample has been placed in one arm of the interferometer, thus, in addition to absorption, causing uncompensated group velocity dispersion, i.e., different frequencies propagate with nonlinearly related phase velocities. Therefore, the IR interferograms appear broadened and asymmetric. Such a raw interferogram for a polypropylene film and a reference interferogram (no sample inserted) are shown in Figs.~\ref{fig:pp_int_sa}~and~\ref{fig:pp_int_ref} respectively.

The result of the Fourier transform in the general case has a complex form. However, the phase spectra representing the frequency-specific delay can be utilized to access the uncompensated dispersion, i.e., the sample is in one of the arms only. More details on the calculation procedure in dispersive or so-called asymmetric FT spectroscopy can be found in the specialized literature\cite{birch_dispersive_1987,doi:10.1080/00107519008213783, chalmers_handbook_2001, Bell1967,DABROWSKA2023122014}. 

The phase is derived from the imaginary and real parts of the Fourier transform spectrum as:
\begin{equation}
\phi(\tilde{\nu}) = \arctan{\left[\frac{\mathrm{Im}\langle\mathfrak{F}(S(\delta))\rangle}{\mathrm{Re}\langle\mathfrak{F}(S(\delta))\rangle}\right]},\label{eq:phase}
\end{equation}
where $S(\delta)$ represents the raw signal that depends on the arm length difference $\delta$. 

\begin{figure}[ht]
         \centering
         \includegraphics[width=1\linewidth]{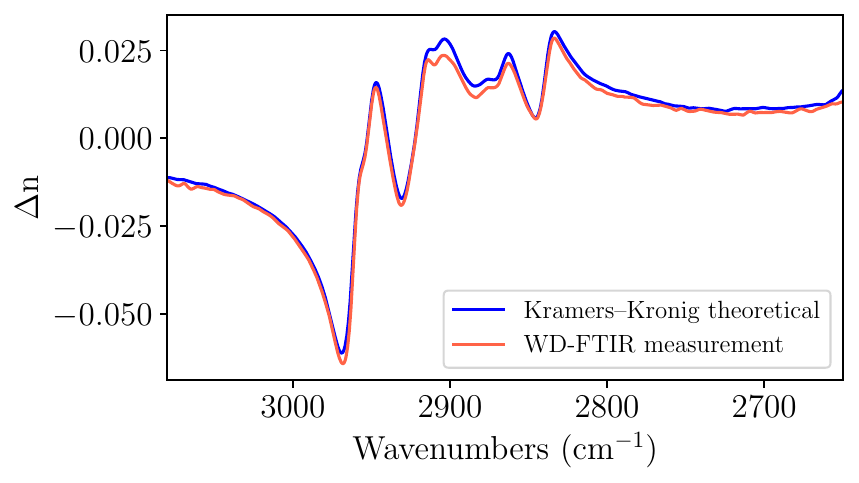}
\caption{\label{fig:pp-disp} Dispersion spectrum of polypropylene (anomalous, near an absorption band) measured and compared to the theory (derived using the Kramers-Kronig dispersion relation).}
\end{figure}

In order to obtain smooth phase spectra, the calculated phase (Eq.~\ref{eq:phase}) is unwrapped using phase unwrapping algorithms (provided by Numpy Python library [see details in~\cite{WDFTgit}]). The reference phase spectrum (corresponds to a linear slope as it induces a shift) is then subtracted from the sample phase spectrum, resulting in a net sample-induced phase difference $\Delta \phi$. Since the thickness of the sample $d$ is known, the refractive index variations can be obtained in the following form:
\begin{equation}
\Delta n (\tilde{\nu}) = -\frac{\Delta \phi(\tilde{\nu})}{2\pi d \tilde{\nu}}.
\end{equation}
The dispersion profile of the sample is shown in Fig.~\ref{fig:pp-disp}.
Polypropylene features an absorption band within the emission window of the source. The absorption has a distinct shape and is caused by symmetric and asymmetric stretching vibrations of C-H\textsubscript{3} and C-H\textsubscript{2} functional groups\cite{Andreassen1999}. Therefore, a particularly interesting region of anomalous dispersion can be observed and evaluated as depicted in Fig.~\ref{fig:pp-disp}. 

The direct dispersive spectroscopy measurements depicted in Fig.~\ref{fig:pp-disp} are verified using the bidirectional Kramers–Kronig relations that connect the real and imaginary parts of the complex refractive index. For the real part of the refractive index, the relation is defined as:
\begin{equation}
\Delta n = \frac{1}{\pi} \mathrm{p.v.} \int\frac{\tilde{\nu'}k(\tilde{\nu'})}{\tilde{\nu'^2}-\tilde{\nu^2}} d\tilde{\nu'},
\end{equation}
where $k(\tilde{\nu})$ is the imaginary part of the refractive index.
Since spectroscopic transmittance measurements yield both the absorption and dispersion properties of the sample, the imaginary part of the refractive index can be directly obtained from the absorption coefficient $\alpha(\tilde{\nu})$ as:
\vspace{2\baselineskip}
\begin{equation}
k(\tilde{\nu}) = \frac{\alpha(\tilde{\nu})}{4\pi\tilde{\nu}},
\end{equation}
where the absorption coefficient can be calculated using the Bouguer-Beer-Lambert law in its exponential form:
\begin{equation}
-\ln{\frac{I(\tilde{\nu})}{I_0(\tilde{\nu})}} = \alpha(\tilde{\nu}) d.
\end{equation}

\begin{figure}[ht]
         \centering
         \includegraphics[width=1\linewidth]{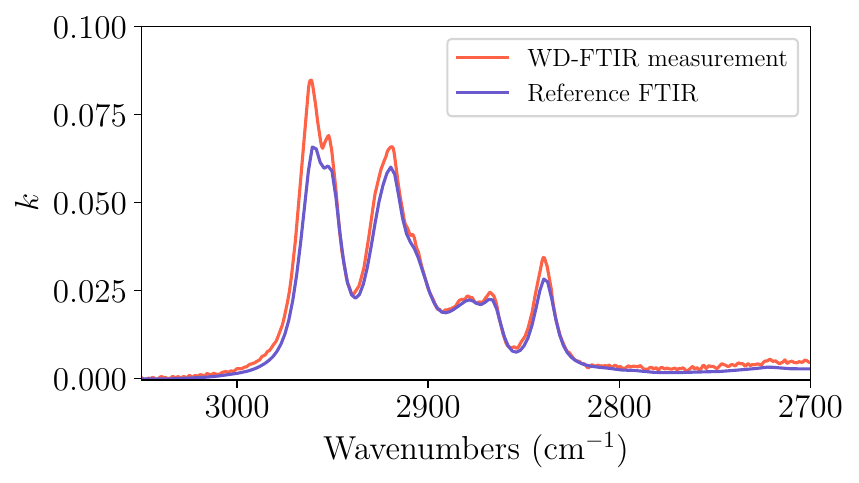}
\caption{\label{fig:pp-k} Imaginary part of the refractive index of polypropylene measured with the developed FT wavefront division spectrometer; the measurement obtained with a conventional state-of-the-art FTIR spectrometer is shown for comparison.}
\end{figure}

Hence, the direct measurement is in remarkably good agreement with the theoretical profile (highlighted in blue in Fig.~\ref{fig:pp-disp}). The calculated imaginary part of the refractive index is shown in Fig.~\ref{fig:pp-k}.
The high quality of the absorption measurement is confirmed by a reference measurement obtained with a standard state-of-the-art commercial FTIR spectrometer (Bruker, Vertex~70, see Fig.~\ref{fig:pp-k}).
\subsection{Reflection measurements: insulin products}

In this section, we demonstrate another measurement mode: reflection measurements. The proposed FT scheme relies on the high spatial coherence of light sources, thus offering inherent advantages for this measurement modality. As the properties are related, sources with high spatial coherence typically exhibit an M\textsuperscript{2} value close to 1 and excellent spectral brightness. This characteristic enables the removal of the common trade-off between spatial resolution and sensitivity found in state-of-the-art FTIR- and micro-spectrometry and based on thermal emitters\cite{Kilgus:18}. Consequently, the same measurements can be accomplished using wavefront division interferometry but with a higher speed and signal-to-noise ratio. Spatial resolution, which depends on the beam quality and optics, can approach or reach the diffraction limit. These factors indicate the high potential of the developed approach for various hyperspectral imaging applications, in particular for chemical mapping and stand-off applications. Here, as a proof-of-concept, we demonstrate a single-point measurement without spatial scanning. The demonstrated measurement is intended to illustrate the possibilities of using the concept in real industrial monitoring.

The reflection measurements were realized in the combined beam scenario. Modifications in one arm of the interferometer necessitate matching the arm length of the other arm. Therefore, the manipulation of the combined beam is more feasible in the case of the reflection geometry. The combined beam was coupled into a single-mode fluoride fiber and connected to a reflection measurements head, where it was collimated using a reflective off-axis parabolic mirror. The high-quality fundamental Gaussian beam of around 12~mm $1/e^2$ diameter was then focused onto the sample using a low-NA ZnSe lens (200~mm). The spot size in the focus is estimated to be around 70~\textmu m. The reflection geometry implied a small angle between the incident and reflected beams, facilitating path separation and eliminating the need for beam splitters. The reflected light was then collected and focused onto the detector using an off-axis parabolic mirror.

The sample selected for analysis was a dried commercial insulin product deposited on a stainless steel substrate.
The sample was quantified to contain approximately 130~\textmu g/cm\textsuperscript{2} of the insulin drug product that comprised: active pharmaceutical ingredient (insulin, main component), glycerol (main excipient), and in smaller concentrations M-cresol, phenol, and sodium phosphate dibasic dihydrate.

\begin{figure}[ht]
         \centering
         \includegraphics[width=1\linewidth]{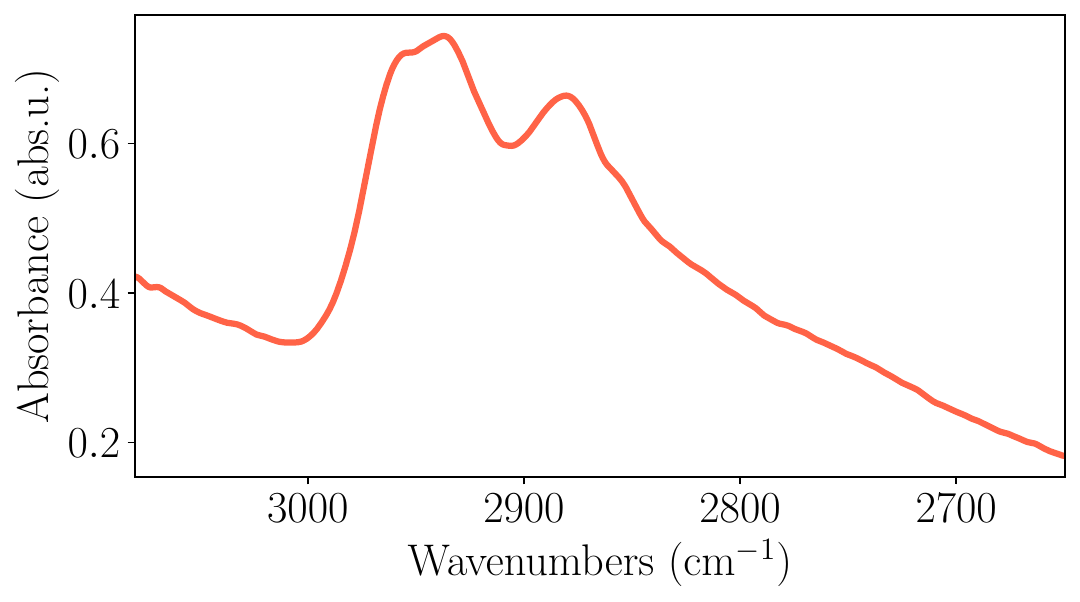}
\caption{\label{fig:insu-abs} Absorbance spectrum of the dried insulin product deposited on a stainless steel substrate measured in reflection geometry (the spot size is around 70~\textmu m).}
\end{figure}

The relevant part of the absorbance spectrum of the compounded insulin formulation, measured in reflection, is displayed in Fig.~\ref{fig:insu-abs}. 
For this chemically complex sample (combining the weighted absorbance of comprising compounds), a very good agreement with publicly available scientific reports was found~\cite{doi:10.1177/1932296820913874}. The observed spectral features can be attributed to C-H stretching vibrations, primarily from methyl (CH\textsubscript{3}) and methylene (CH\textsubscript{2}) groups. 

\section{Discussion on wavefront division interferometry in IR spectroscopy}\label{sec:history}

Interferometric wavefront division schemes are well-known as they have played a major role in fundamental research. Nowadays, these techniques represent an established academic field and have a prominent place in the history of science. Early wavefront division interferometric experiments encompass well-known implementations such as Young's double slit arrangement, Billet's split lens system, or Fresnel's double mirror interferometer. 
The idea of interfering fronts of a single separated wave has also been used, for example, in optical coherence tomography\cite{Podoleanu:07,ZHANG20121589, 10.1117/12.741098} and dispersion-free delay lines for measurements of ultrashort laser pulses\cite{mashiko_all-reflective_2003, tyagi_attosecond_2022}. 
An extensive literature review shows that various wavefront division interferometric schemes were also adopted and applied for broadband spectrometry, including early designs at the dawn of the development of FT spectroscopy\cite{Richards:64, Strong:57}. These include a wide range of implementations primarily from lamellar grating interferomers\cite{Strong:60,Manzardo:04} and Mach-Zehnder configurations with two wavefront-dividing reflection-transmission beam splitters\cite{YIN2000529} to specific variations of Fresnel bi-mirror scanning FT systems\cite{de_oliveira_high-resolution_2011,10.1063/1.3111452}. The aforementioned grating interferometers use a system of several alternating strip mirrors placed at different heights to produce a two-beam (reflected from each of the two mirrors seen at zero order) interference. The wavefront division Mach-Zehnder variation employs lossless transmission-reflection wavefront beamsplitters (formed by aluminium-coated reflecting strips on 1~mm thick transmission glass plates), while the principle of operation is similar to lamellar grating interferometers. 
The FT bi-mirror scanning systems, in turn, utilize the observation of an interference pattern in the local region of wave overlap induced by a tilted mirror arrangement. 

These early wavefront division systems have proven their applicability and laid a solid foundation for real broadband wavefront division FT spectrometry. However, they did not become widespread and common, yielding the first position for routine laboratory use to Michelson-type interferometer variations. The potential reasons for this are technical complexity and specific limitations. Grating interferometers, despite their broad achromaticity, are complex systems, require the fabrication of sophisticated grating mirrors, and have strong diffraction losses (zeroth diffraction order is used to detect temporal interferograms).
The coherent region of incident light in grating interferometers must be at least twice as wide as the grating period, which limits their application and establishes a certain compromise between diffraction losses (high stripe density) and coherent properties of the source. In the bi-mirror FT spectrometers, the interfering beams are not collinear, which leads to alignment problems, scanning range limitations, and the development of a linear pattern of numerous fringes in the detector plane and, therefore, effectively restricting Jacquinot's advantage\cite{10.1063/1.3111452}.

The wavefront division interferometer concept proposed here is designed to bridge the advantages of both amplitude and existing wavefront division concepts relying on properties of novel laser sources: losses are minimized by using non-diffractive wavefront division schemes; extreme bandwidth is supported by all-mirror components; brightness of lasers is effectively exploited. The demonstrated practical implementation has a simple and robust design. 
Thus, the scheme proposed here is characterized by simplicity, efficiency and flexibility. 
Our arrangement is generally source-agnostic provided that the source has high spatial coherence (extended thermal sources can be adapted through spatial filtering). Hence, the concept is specifically designed to leverage and fully exploit novel tabletop laser sources with high spatial coherence, exemplified here for mid-IR supercontinuum lasers. 
The novel mid-IR sources have been extensively matured in the last decade, and to the best of our knowledge, their use in FT wavefront division spectrometry has not been previously demonstrated. We anticipate following up research on potential studies with extended broadband emitters across multiple spectral ranges. 
Combined with these emitters, our solution promises to provide a compact and robust spectroscopic instrument with high spectral and spatial performance due to orders of magnitude higher brightness\cite{Zorin:22} and ultra-broad coverage. Precision, stability, dynamic range optimization, and other system enhancements are engineering challenges, with no physical limitations anticipated beyond those present in standard FTIR instruments.

\section{Conclusion and outlook}\label{sec:end}

We have presented a novel, elegant concept of a wavefront division interferometer for broadband spectrometry relying on novel sources with high spatial coherence. The system can potentially operate from the ultraviolet to terahertz range. The theoretical framework governing the interference pattern has been provided and verified through optical propagation simulations and experiments. 

Based on the theoretical foundations, an optical scheme based on standard off-the-shelf components for an all-mirror infrared (IR) Fourier transform spectrometer has been proposed. The all-mirror design stands out in its simplicity while offering several advantages over state-of-the-art systems, such as ultra-broad achromaticity and dispersion-free operation. In addition, the spectrometer possesses high throughput and the capability of dynamically adjusting the split ratio between the interferometric arms. The system's capabilities and performances have been demonstrated for various routine spectroscopic measurements; the high quality of absorption and dispersion spectra has been showcased. The operation of the system has been demonstrated by transmittance measurements of methane and polypropylene and by reflectance measurements of dried insulin products.

In the experimental part, a novel mid-IR supercontinuum source was used for pilot validation and as a proof-of-concept. The employed source has a broad spectral coverage (1.1\textmu m - 4.4~\textmu m). However, we envisage great system capabilities for supercontinuum sources with much wider spectral coverage\cite{Martinez:s,OB133,16um,Zhang:19}, as well as for a range of other broadband laser systems including (but not limited to) Fabry–Pérot quantum cascade lasers or nonlinear systems\textemdash e.g., those based on standard parametric architectures or remarkable novel intrapulse difference frequency generation schemes\cite{Krebbers:24, elu_seven-octave_2021}. Such high spatial coherence broadband sources could utilize the full capabilities of the proposed interferometric system. In view of the ultra-broadband spectral response of the detector used, combined with the overall achromaticity of the system, the spectroscopic measurements, therefore, can be performed over extremely wide spectral ranges. This is of great interest in, for instance, mid-IR spectroscopy and microscopy.

The careful reader may notice the possibility of miniaturization and simplification of the proposed spectrometer architecture. The wavefront dividing and combining prisms and the interferometer mirrors (M1 with M3 and M2 with M4) can be united into a metal-coated cube and two retroreflectors, respectively. This kind of modified system represents a compact solution with a 4-fold enhancement in the scan length. In addition, the use of pre-manufactured retroreflectors and cubes guarantees the stability of the optical scheme. Therefore, the use of similar modifications is anticipated in future due to their relevance for industrial and laboratory applications. This will support and stimulate further development of wavefront division Fourier transform spectroscopy techniques. 

\begin{acknowledgments}
Ivan Zorin is grateful to Andrii Prylepa for early discussions over a cup of coffee about the possibility of transferring the concept of wavefront division from OCT to spectrometry.

The authors would like to thank David Wimberger for preparing the insulin samples.

The authors acknowledge financial support from the European Union's Horizon Europe Research And Innovation Programme Under Grant Agreement No.101057844 (Enviromed project).

This project is co-financed by research subsidies granted by the government of Upper Austria: Grant Nr. FTI 2022 (HIQUAMP): Wi-2021-303205/13-Au.
\end{acknowledgments}

\section*{Data Availability Statement}
The data that support the findings of this study are available from the corresponding author upon reasonable request.

\bibliography{main}
\end{document}